# Unraveling the Geometry of the Crab Nebula's "Inner Ring"

Martin C. Weisskopf, Ronald F. Elsner, Jeffery J. Kolodziejczak, Stephen L. O'Dell, and Allyn F. Tennant

Space Science Office, NASA Marshall Space Flight Center, Huntsville, AL 35812

Abstract

*Chandra* images of the Crab Nebula resolve the detailed structure of its "inner ring", possibly a termination shock where pulsar-accelerated relativistic particles begin to emit X radiation. Analysis of these images finds that the center of the ellipse—presumably a circular ring in projection—lies about 0.9″ (10 light-days at 2 kpc) from the pulsar's image, at a position angle of about 300° (East of North). This analysis also measures properties of the ellipse: The position angle of the semi-major axis is about 210° (East of North); the aspect ratio, 0.49.

In a simple—albeit, not unique—de-projection of the observed geometry, a circular ring is centered on the axis of symmetry of the pulsar wind nebula. This ring is not equatorial but rather lies near +4.5° latitude in pulsar-centered coordinates. Alternative geometries are briefly discussed.

## 1. INTRODUCTION

Images obtained with the *Chandra X-ray Observatory* reveal the rich and dynamic x-ray structure of the Crab pulsar wind nebula (Gaensler & Slane 2006, and references therein). A persistent feature in these images is the "inner ring", posited to be the termination shock of the pulsar's relativistic wind (Rees & Gunn 1974; Kennel & Coroniti 1984). Close inspection of *Chandra* data (Figure 1) finds that the inner ellipse—presumably the projection of a circular ring—appears *not* to be centered on the pulsar. In this paper we quantify this de-center in the image and seek to account for it in terms of a simple, de-projected geometric model.

First (§2) we describe the observations and data reduction. Next (§3) we discuss the analysis of the image and measurement of the geometric properties of the inner ellipse. Then (§4) we present a simple (de-projected) geometric model. Finally (§5) we briefly consider alternative models.

## 2. OBSERVATIONS AND DATA REDUCTION

For this study of the Crab Nebula, we analyzed the deepest and spatially most precise *Chandra* zeroth-order images obtained using the Low-Energy Transmission Grating (LETG) with the High-Resolution Camera Spectroscopy (HRC-S) detector: ObsID 758 is a 100-ks telemetry-saturated observation on 2000 January 31; ObsID 759 is a 50-ks telemetry-saturated observation on 2000 February 2. Due to its shorter integration time and increased telemetry saturation (resulting from a difference in the HRC-S configuration), the second observation provides only about 20% of the number of events collected in the first observation. We processed the data from the *Chandra* Archive, using CIAO 4.0 (Fruscione 2006) with CalDB 3.4.0.

Figure 1 shows a LETG/HRC-S zeroth-order image for the merged data (ObsIDs 758 and 759). A prominent feature in the x-ray image is the "inner ring"—an ellipse, presumably a circular structure in projection—that was apparent from the first *Chandra* images of the Crab Nebula (Weisskopf et al. 2000). Close inspection of this image reveals that the inner ellipse is *not* centered on the pulsar.

### 3. ANALYSIS AND RESULTS

Our analysis requires measuring the geometry of the brightened elliptical inner "ring". To do this, we generated 36 radial profiles of surface brightness, one for each 10° position-angle sector centered on the pulsar. Figures 2 and 3 show these radial profiles, between 10 and 120 HRC pixels from the pulsar, where one HRC (electronic) pixel is 6.43 μm, corresponding to 0.1318".

To establish a radial distance from the pulsar to the "ring" structure we first visually inspected each radial profile (Figures 2 and 3). For a few position-angle bins, we do not see a well-defined feature in the profile and must omit those bins from the analysis. We also omitted the 4 data values between position angles 100° and 140°, where the Crab jet crosses the "inner ring". Using the remaining radial profiles, we use the local peak surface brightness in each bin to characterize the inner ellipse. Alternatively, we fit a hyperbolic tangent to a portion of the radial profile to measure the radius to the inner edge of this ellipse for each position angle. Table 1 lists the measured radius $r_p$ for the peak flux and radius $r_i$ for the inner edge for the 36 bins in position angle $\varphi$. For $r_i$, Table 1 also quotes a statistical error for the position of the inner edge, based upon the fit of a hyperbolic tangent to the radial profile.[1]

Table 2 characterizes the ellipse, with best-fit parameters and corresponding uncertainties determined by a least-squares fit[2] to the measured distribution $r_p(\varphi)$ or $r_i(\varphi)$. The derived characteristics of the x-ray ellipse (Table 2) are similar to previously determined values (e.g., Weisskopf et al. 2000). The important new result is that the center of the ellipse lies about 1.0" (about 11.5 light days at 2 kpc) from the projected position of the pulsar. Furthermore, the direction of this offset is consistent with it lying along the minor axis of the ellipse—i.e., perpendicular to the major axis. This leads us to suggest a simple model for the de-projected geometry.

### 4. DE-PROJECTED GEOMETRY

In this simple geometric model (Figure 4), a circular ring of radius R is centered on the axis of symmetry of the pulsar wind nebula (PWN)—presumably the pulsar's spin axis. The line of sight to the observer is at a polar angle $\theta$ to the axis of symmetry, which is projected at a position angle $\psi$ (East of North) on the sky. By convention, we take the PWN's positive axis of symmetry (z) to lie in the northwest quadrant, such that $\psi \approx 300°$. To resolve the ambiguity in the polar angle $\theta$, we interpret the brighter regions of the Crab torus (outside the "inner ring") to result from relativistic enhancement of emission from outflowing material (Pelling et al. 1987). This then implies that the brighter regions of the torus are nearer the observer, which requires that $\theta > 90°$—i.e., we are observing the "bottom" of the ring.

---

[1] We have not assigned a statistical error to $r_p$ because we simply selected the brightest local pixel rather than fitting a function to locate the peak. We deemed this simpler approach acceptable, in that other errors dominate the statistical errors. (See next footnote.)

[2] Actually, we performed the fitting by minimizing $\chi^2$. As the fits were not statistically acceptable, we added (in quadrature) "systematic" errors to the statistical errors, in order to raise the $\chi^2$ to 1 per degree of freedom. The resulting uncertainties in the fitting parameters are approximately the same as would be obtained in a least-squares analysis.

Up to this point, the model is essentially the same as that used previously to de-project the x-ray inner ring (e.g., Weisskopf et al. 2000; Ng and Romani 2004, 2008), the x-ray torus (Aschenbach & Brinkmann 1975), and visible-light structure (Hester 2002, 2008). Various magnetohydrodynamic calculations and simulations (e.g., Kennel & Coroniti 1984; Bogovalov & Khangoulian 2002; Del Zanna, Amato, & Bucciantini 2004; Kirk, Lyubarsky, & Petri 2009 and references therein) yield a torus–jet structure resulting from an equatorial striped wind. Here, however, the ring does not lie exactly in the equatorial plane of the pulsar, but rather at low (but non-zero) latitude $\lambda$.

Table 3 gives the best-fit parameters with their respective 1-sigma uncertainties, as determined by a least-squares fit of the projected model to the measured distribution $r_p(\varphi)$ or $r_i(\varphi)$. Figure 5 shows the fit of the model to the measured distribution $r_i(\varphi)$ of the inner edge of the ellipse. In the context of this simple geometric model, the key result is that the inner ring is at a latitude $\lambda \approx +4.5°$ in the pulsar-centered coordinate system.

## 5. Discussion and Conclusions

The simple geometric model described above is not a unique de-projection. Here we briefly discuss alternative models that might account for an apparent oval structure that is not centered on the pulsar.

First we consider geometrical models that, like the above model, assume an azimuthally symmetric physical structure—namely, a circular ring aligned with and centered on the pulsar's spin axis. As we have seen, a non-equatorial ring can account for a de-center of the projected ellipse from the pulsar. A centered, circular, equatorial ring could be consistent with the image under at least two conditions.

One possibility is that the inner ring, which appears to be a toroidal structure a couple arcseconds thick, contains a mildly relativistic inward flow of material from the shock in addition to the mildly relativistic outflow. If this were the case, then the x-ray brightness distribution would not be azimuthally symmetric even if the density of emitting material were. The sense of this effect agrees with the observed image, in that the x-ray brightness from the more distant half of the torus would appear closer to the pulsar than that from the nearer half.

A second possibility is that the radius of the termination shock changes slightly with time. During epochs of increasing radius, retardation effects would result in the more distant half of the torus appearing closer to the pulsar than the nearer half. The time delay between the far and near regions of the inner ring is about 260 days, thus this model would require an approximately 20 light-day increase in radius over this period, in order to account for the observed de-center. Note that such a model can account for the observed de-center only episodically, during epochs when the shock's radius increases: The sign of this effect is opposite during epochs when the shock's radius decreases. Thus, it is subject to verification. Indeed, examination of about 20 *Chandra* ACIS images, obtained over a decade, finds the southeast sector of the ring to be closer than the northwest sector in every case. This finding then tends not to support such an explanation.

A third possibility is that the ring is not physically centered on the pulsar. This would occur if the radius of the termination shock depended upon azimuth, due to a gradient in the ambient pressure. In contrast with the models discussed thus far, the direction of the resulting de-center need not lie along the minor axis of the observed ellipse.

Finally, we note that some axisymmetric equatorial-flow simulations yield more complex patterns in the observed intensity, due to caustics resulting from relativistic beaming. For example, Figures 10 and 11 of Komissarov & Lyubarsky (2004) predict two unequal arcs (not a full or symmetric ring), each offset from the pulsar by differing projected distances. Such a pattern is possibly consistent with the observed x-ray images of the "inner ring".

In summary, *Chandra* images of the Crab pulsar wind nebula show that the center of the inner ellipse is offset from the pulsar by about 0.9". While the simple geometric model of a non-equatorial ring can account for the observed displacement, other de-projections are possible. More detailed theoretical studies of the structure of pulsar wind nebulae need to take into account the observed offset.

| Table 1. Radius to feature in surface brightness |||||||
|---|---|---|---|---|---|
| Angle[1] range (degrees) | Peak $r_p$ (HRC pixels) | Inner edge $r_i$ (HRC pixels) | Angle[1] range (degrees) | Peak $r_p$ (HRC pixels) | Inner edge $r_i$ (HRC pixels) |
| 0-10 | 83.8 | $76.7 \pm 1.3$ | 180-190 | 76.3 | $62.1 \pm 6.9$ |
| 10-20 | 98.8 | $84.4 \pm 1.2$ | 190-200 | 86.3 | $54.5 \pm 8.1$ |
| 20-30 | 96.3 | $87.4 \pm 1.3$ | 200-210 | 98.8 | $81.7 \pm 15.0$ |
| 30-40 | 96.3 | $86.2 \pm 2.5$ | 210-220 | 108.8 | $90.2 \pm 2.1$ |
| 40-50 | 88.8 | $75.6 \pm 14.3$ | 220-230 | 86.3 | $76.3 \pm 1.6$ |
| 50-60 |  | $45.2 \pm 15.1$ | 230-240 | 86.3 | $75.0 \pm 1.7$ |
| 60-70 | 53.8 | $47.9 \pm 0.74$ | 240-250 | 78.8 | $62.7 \pm 4.2$ |
| 70-80 | 53.8 | $48.7 \pm 1.1$ | 250-260 | 68.8 | $61.4 \pm 0.7$ |
| 80-90 | 46.3 | $43.7 \pm 0.3$ | 260-270 | 63.8 | $58.3 \pm 0.8$ |
| 90-100 | 31.3 |  | 270-280 | 63.8 | $55.8 \pm 0.8$ |
| 100-110 | 36.3 | $25.5 \pm 1.3$ | 280-290 | 63.8 | $52.2 \pm 0.7$ |
| 110-120 | 36.3 | $30.5 \pm 1.3$ | 290-300 | 61.3 | $49.9 \pm 0.7$ |
| 120-130 | 36.3 | $30.1 \pm 0.8$ | 300-310 | 53.8 | $48.3 \pm 0.9$ |
| 130-140 | 41.3 | $27.0 \pm 3.2$ | 310-320 | 53.8 | $48.7 \pm 0.7$ |
| 140-150 | 41.3 |  | 320-330 | 61.3 | $52.5 \pm 0.9$ |
| 150-160 | 43.8 |  | 330-340 | 63.8 | $55.1 \pm 0.7$ |
| 160-170 | 58.8 | $51.1 \pm 1.6$ | 340-350 | 71.3 | $58.8 \pm 1.1$ |
| 170-180 | 66.3 |  | 350-360 |  | $68.6 \pm 0.8$ |
| [1] Angles measured East of North ||||||

| Table 2. Geometric properties of the inner ellipse (2D) |||
|---|---|---|
| Feature characterizing the ellipse | Peak brightness | Inner edge |
| Semi-major axis, a | $13.31'' \pm 0.23''$ | $11.30'' \pm 0.21''$ |
| Semi-minor axis, b | $6.44'' \pm 0.13''$ | $5.64'' \pm 0.11''$ |
| Aspect ratio, (b/a) | $0.484 \pm 0.013$ | $0.499 \pm 0.014$ |
| Angular eccentricity, arccosine(b/a) | $60.1° \pm 0.9°$ | $61.1° \pm 0.9°$ |
| Elongation direction (E of N) | $209.0° \pm 1.1°$ | $209.6° \pm 1.4°$ |
| Direction of center from pulsar (E of N) | $289° \pm 11°$ | $309° \pm 13°$ |
| Distance of center from pulsar, $\xi_\perp$ | $0.95'' \pm 0.11''$ | $0.82'' \pm 0.11''$ |

| Table 3. Geometric properties of a simple de-projected model (3D) |||
|---|---|---|
| Feature characterizing the inner ring | Peak brightness | Inner edge |
| Radius of the inner-ring feature, R | $13.25'' \pm 0.26''$ | $11.38'' \pm 0.23''$ |
| Polar angle from axis to line of sight, $\theta$ | $119.1° \pm 1.0°$ | $119.6° \pm 1.0°$ |
| Position angle of projected axis (E of N), $\psi$ | $298.4° \pm 1.1°$ | $298.7° \pm 1.2°$ |
| Axial distance to plane of the inner ring, $\xi$ | $1.04'' \pm 0.13''$ | $0.91'' \pm 0.11''$ |
| Latitude of the inner ring, $\lambda$ | $+4.49° \pm 0.77°$ | $+4.57° \pm 0.77°$ |

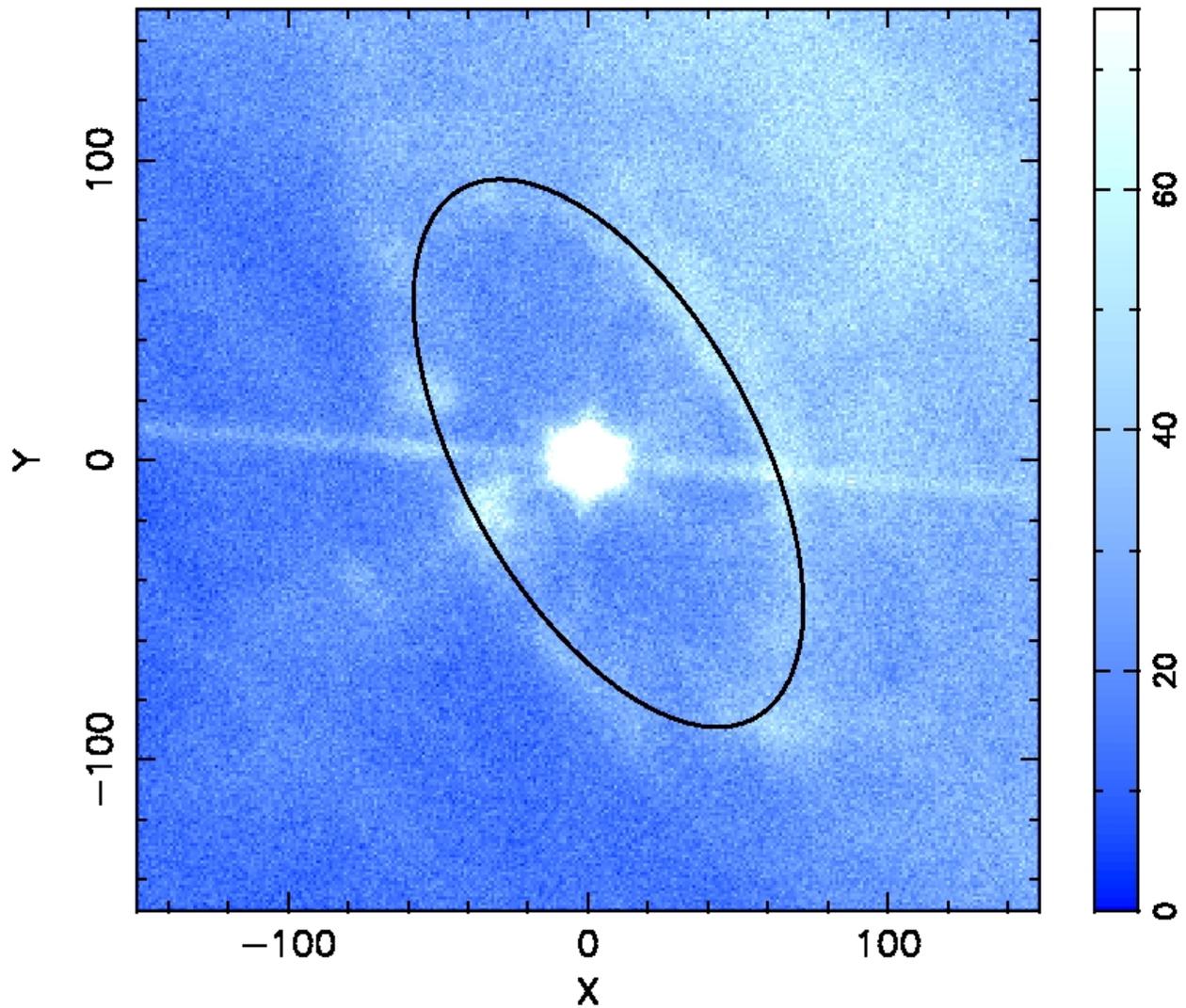

Figure 1. *Chandra* LETG/HRC-S zeroth-order image of the Crab Nebula (ObsIDs 758 and 759), over a 40″×40″ field centered on the pulsar. The oval line marks the peak brightness of the ellipse, which is *not* centered on the pulsar. North is along the +Y axis; East, along the –X axis. The units are HRC pixels (0.132″). The approximately E-W streak results from diffraction (in the cross-dispersion direction) by support bars in the LETG facets.

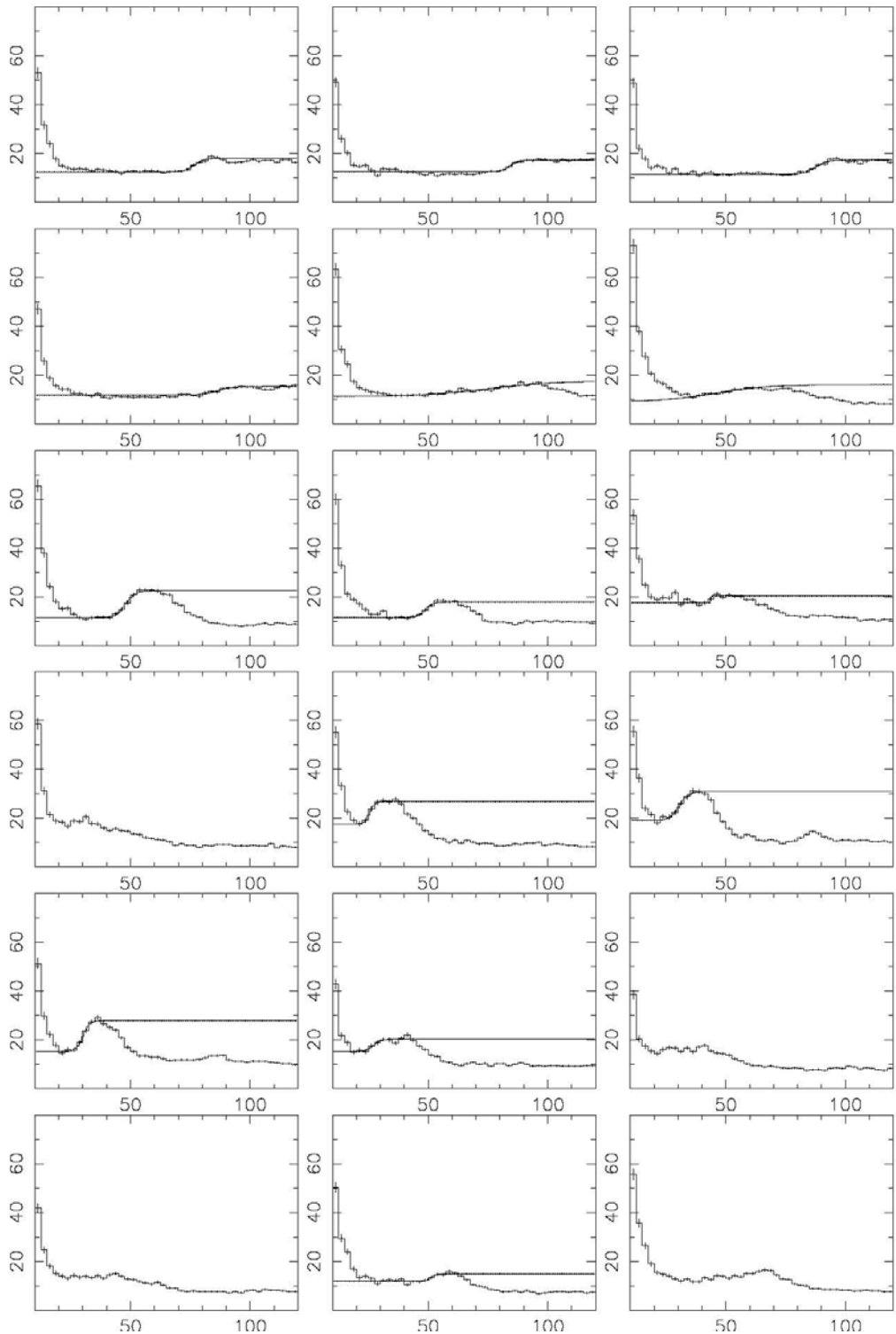

Figure 2. Surface brightness versus radius (in HRC pixels) from the pulsar. The images, from the top left to the bottom right, represent successive $\Delta\varphi = 10°$ slices over $0° < \varphi \leq 180°$ East of North. The smooth solid lines are fits of a hyperbolic tangent to the inner edge of the bright elliptical ring.

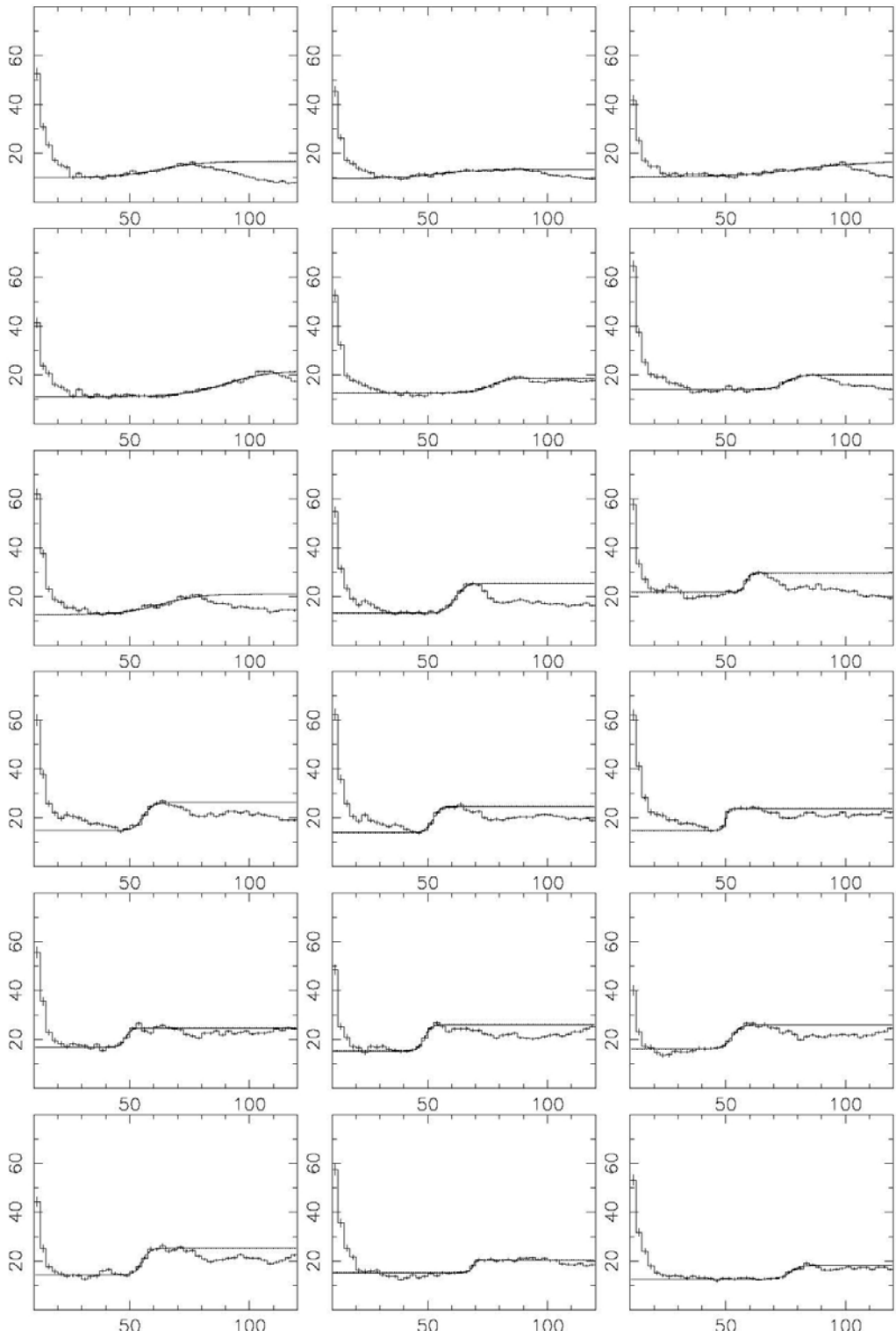

Figure 3. Surface brightness versus radius (in HRC pixels) from the pulsar. The images, from the top left to the bottom right, represent successive $\Delta\varphi = 10°$ slices over $180° < \varphi \leq 360°$ East of North.

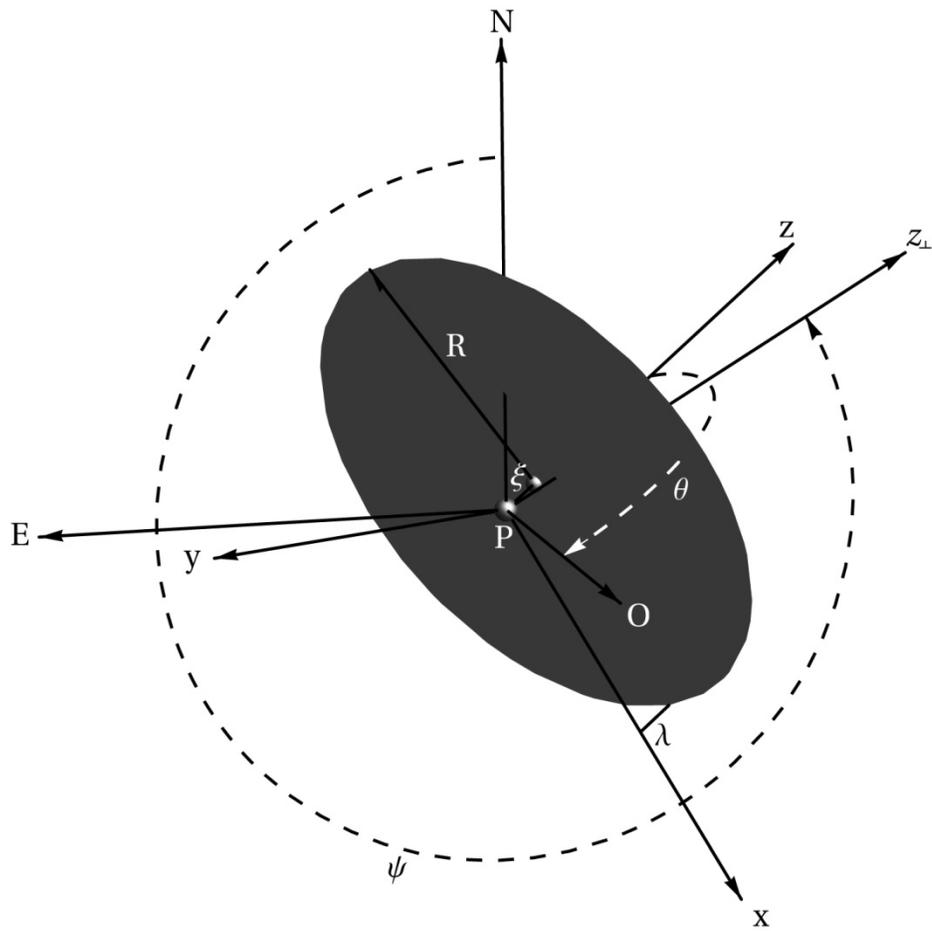

Figure 4. Possible de-projected geometry. A ring of radius R is oriented to and centered on the symmetry axis z of the pulsar wind nebula. However, the ring lies in a plane displaced an axial distance $\xi$ from the pulsar "P", corresponding to a latitude $\lambda = \tan^{-1}(\xi/R)$. The projection $z_\perp$ of the symmetry axis onto the sky lies at an angle $\psi$ East of North. The direction to the observer "O" is at an angle $\theta$ from the pulsar's axis of symmetry z.

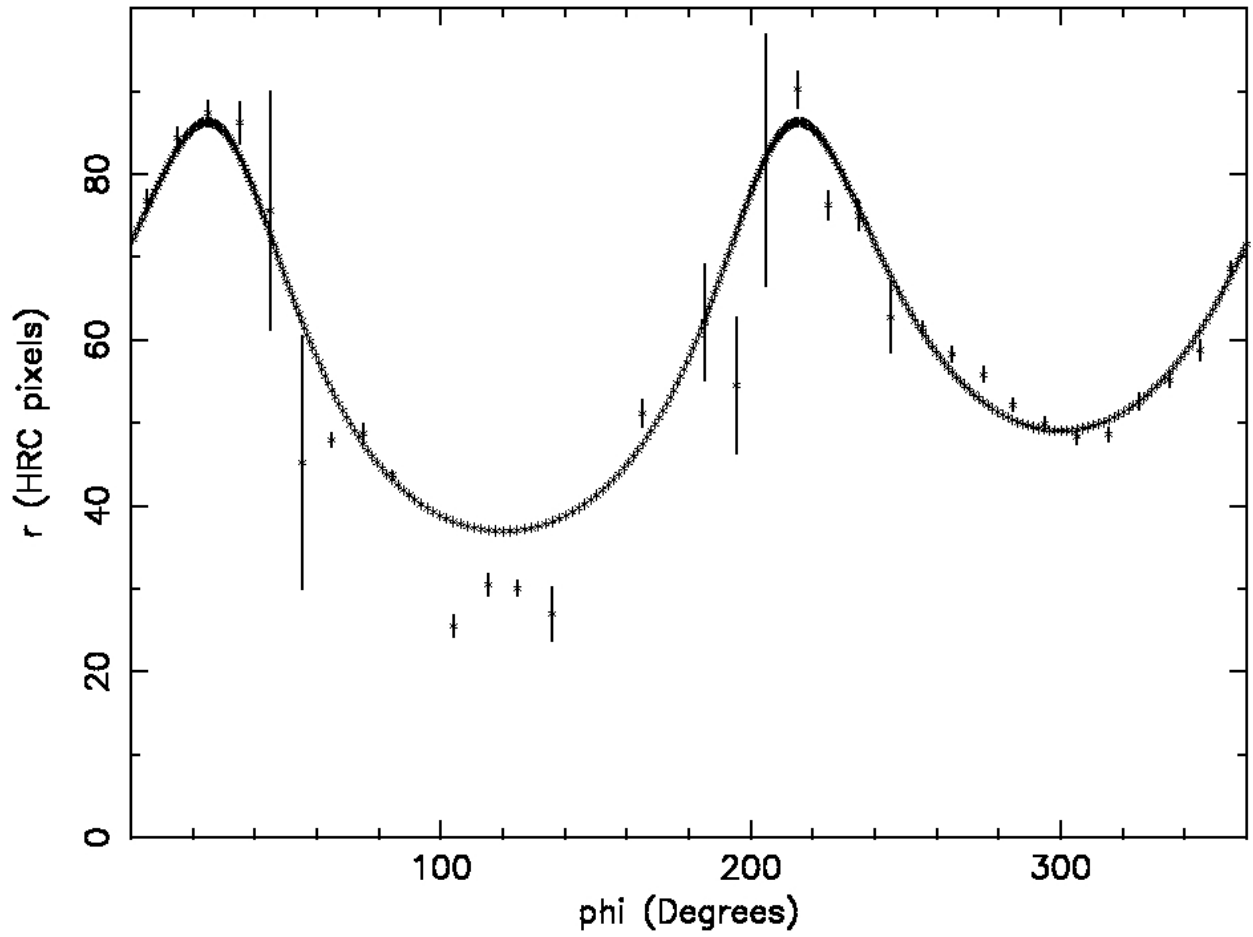

Figure 5. Radius from the pulsar of the inner edge of the "ring" as a function of angle (East of North). The markers with (statistical) error bars are the $r_i(\varphi)$ data (Table 1); the line is the best-fit model (Table 3) to these data. Note that the fit ignores the 4 data points in vicinity of the southern jet, between 100° and 140°.